\documentclass[DIV24,twocolumn,9pt]{scrartcl}
\pdfoutput=1
    
    \usepackage{ifxetex}
\ifxetex
\usepackage{fontspec}
\else
\usepackage[T1]{fontenc}
\usepackage[utf8]{inputenc}
\fi
\usepackage{textcomp}
\usepackage{graphicx}
\usepackage{times}
\usepackage{hyperref}

\usepackage{fixltx2e}
\usepackage{amssymb}
\usepackage{fancyhdr}

\usepackage{color}

\usepackage{dblfloatfix}

\usepackage[font={rm},labelfont={rm,bf},small,]{caption}
\setcapindent{0cm}
\DeclareCaptionLabelSeparator{pp}{ | }

\DeclareCaptionType{SI-figure}[Fig.]
\DeclareCaptionLabelFormat{bf-parens-si}{\bfseries Extended Data Figure #2 }
\DeclareCaptionLabelFormat{bf-parens-table}{\bfseries Extended Data Table #2}
\definecolor{color06}{rgb}{1.00,0.00,0.00}
\definecolor{color21}{rgb}{0.13,0.13,0.13}
\definecolor{color22}{rgb}{0.44,0.19,0.63}
\definecolor{color23}{rgb}{0.13,0.13,0.13}

\addtokomafont{pagenumber}{\small\sffamily}

\usepackage{url}
\begin{document}

\twocolumn[{
    \textbf{\textsf{This manuscript is the pre-submission manuscript provided by the authors.\\ For the final, post-review version, please see:\\ \url{http://www.nature.com/nature/journal/vaap/ncurrent/full/nature24647.html}}}

    \bigskip

\hrule width \hsize \kern 0.5mm 
\hrule width \hsize \kern 0.5mm 
\hrule width \hsize height 0.5mm

\begin{@twocolumnfalse}

\begin{flushleft}
    \fontsize{23}{8}\selectfont
  \bfseries Discovery of a big void in Khufu's Pyramid by observation of cosmic-ray muons
  \end{flushleft}
  \bfseries
  Kunihiro Morishima\textsuperscript{1}, Mitsuaki Kuno\textsuperscript{1}, Akira 
  Nishio\textsuperscript{1}, Nobuko Kitagawa\textsuperscript{1}, Yuta Manabe\textsuperscript{1},\textsuperscript{ 
  }Masaki Moto\textsuperscript{1}
  --- 
 Fumihiko Takasaki\textsuperscript{2}, Hirofumi Fujii\textsuperscript{2}, Kotaro Satoh\textsuperscript{2}, Hideyo Kodama\textsuperscript{2}, 
  Kohei Hayashi\textsuperscript{2}, Shigeru Odaka\textsuperscript{2} --- Sébastien Procureur\textsuperscript{3}, David Attié\textsuperscript{3}, Simon Bouteille\textsuperscript{3}, 
  Denis Calvet\textsuperscript{3}, Christopher Filosa\textsuperscript{3}, Patrick 
  Magnier\textsuperscript{3}, Irakli Mandjavidze\textsuperscript{3}, Marc Riallot\textsuperscript{3}
  ---
   Benoit Marini\textsuperscript{5}, Pierre Gable\textsuperscript{7}, Yoshikatsu Date\textsuperscript{8}, 
  Makiko Sugiura\textsuperscript{9}, Yasser Elshayeb\textsuperscript{4}, Tamer Elnady\textsuperscript{10}, 
  Mustapha Ezzy\textsuperscript{4}, Emmanuel Guerriero\textsuperscript{7}, Vincent 
  Steiger\textsuperscript{5}, Nicolas Serikoff\textsuperscript{5},\textsuperscript{ 
  }Jean-Baptiste Mouret\textsuperscript{11}, Bernard Charlès\textsuperscript{6}, 
  Hany Helal\textsuperscript{4,5},  Mehdi Tayoubi\textsuperscript{5,6}
  
\bigskip  
  Corresponding Authors: Kunihiro Morishima (\href{mailto:morishima@flab.phys.nagoya-u.ac.jp}{morishima@flab.phys.nagoya-u.ac.jp}) and Mehdi Tayoubi (\href{mailto:tayoubi@hip.institute}{tayoubi@hip.institute})
  	\bigskip

      \normalfont

\textsuperscript{\textit{1}}\textit{F-lab, Nagoya University. Furo-cho, Chikusa-ku, Nagoya, 464-8602, Japan}
--- 
\textsuperscript{\textit{2}}\textit{KEK,}\textit{1-1 
oho, Tsukuba, Ibaraki 305-0801 Japan}
---
\textsuperscript{{\textit{3}}}{\textit{ IRFU,~CEA, 
Université Paris Saclay, 91191 Gif-sur-Yvette, France}}
---
\textsuperscript{{\textit{4}}}{\color{color21} \textit{Cairo University,~9 
Al Gameya, Oula, Giza Governorate, Egypt}}
---
\textsuperscript{{\textit{5}}}{\textit{HIP Institute,~50 
rue de Rome 75008 Paris, France}}
---
\textsuperscript{{\textit{6}}}{\textit{Dassault 
Systèmes,~10 Rue Marcel Dassault, 78140 Vélizy-Villacoublay, France}}
---
\textsuperscript{{\textit{7}}}{\textit{ Emissive,~71 
rue de Provence~75009 Paris, France}}
---
\textsuperscript{{\textit{8}}}{ \textit{NEP,~4-14 
Kamiyama-cho, Shibuya-ku,Tokyo 150-0047, Japan}}
---
\textsuperscript{{\textit{9}}}{\textit{Suave images, 
N-2 Maison de Shino, 3-30-8 Kamineguro, Meguro-Ku, Tokyo, 153-0051, Japan}}
---
\textsuperscript{\textit{10}}\textit{Ain Shams University, Kasr el-Zaafaran, Abbasiya, Cairo, Egypt}
---
\textsuperscript{\textit{11}}\textit{Inria Nancy - Grand Est, 615 rue du Jardin Botanique, Villers-lès-Nancy, 54600, France}

\bigskip
      \normalfont
  \end{@twocolumnfalse}
  }]

\bfseries
\noindent{}The Great Pyramid or Khufu's Pyramid was built on the Giza Plateau (Egypt) during the IV\textsuperscript{th} 
dynasty by the pharaoh Khufu (Cheops), who reigned from 2509 to 2483 BC\textsuperscript{1}. 
Despite being one of the oldest and largest monuments on Earth, there is no consensus 
about how it was built\textsuperscript{2,3}. To better understand 
its internal structure, we imaged the pyramid using muons, which are by-products 
of cosmic rays that are only partially absorbed by stone\textsuperscript{4,5,6}. 
The resulting cosmic-ray muon radiography allows 
us to visualize the known and potentially unknown voids in the pyramid in a non-invasive 
way. Here we report the discovery of a large void (with a cross section similar 
to the Grand Gallery and a length of 30 m minimum) above the Grand Gallery, which 
constitutes the first major inner structure found in the Great Pyramid since the 
19\textsuperscript{th} century\textsuperscript{1}. 
This void, named ScanPyramids Big Void, was first observed with nuclear emulsion 
films\textsuperscript{7,8,9} installed in the Queen's chamber 
(Nagoya University), then confirmed with scintillator hodoscopes\textsuperscript{10,11} 
set up in the same chamber (KEK) and re-confirmed with gas detectors\textsuperscript{12} 
outside of the pyramid (CEA). This large void has therefore been detected with 
a high confidence by three different muon detection technologies and three independent 
analyses. These results constitute a breakthrough for the understanding of Khufu's 
Pyramid and its internal structure. While there is currently no information about 
the role of this void, these findings show how modern particle physics can shed 
new light on the world's archaeological heritage.
\normalfont

The pyramid of Khufu is 139 m high and 230 m wide\textsuperscript{1,13}. There 
are three known chambers (Fig. 1), at different heights of the pyramid, which all 
lie in the north-south vertical plane\textsuperscript{1}: the subterranean chamber, 
the Queen's chamber, and the King's chamber. These chambers are connected by several 
corridors, the most notable one being the Grand Gallery (8.6 m high × 46.7 m long 
× 2.1 to 1.0 m wide). The Queen's chamber and the King's chamber possess two ``air 
shafts'' each, which were mapped by a series of robots\textsuperscript{14,15} between 
1990 and 2010. The original entrance is believed to be the ``descending corridor'', 
which starts from the North face, but today tourists enter the pyramid via a tunnel 
attributed to Caliph al-Ma'mun's (around AD 820)\textsuperscript{1.}

\begin{figure*}[t]
    \includegraphics[]{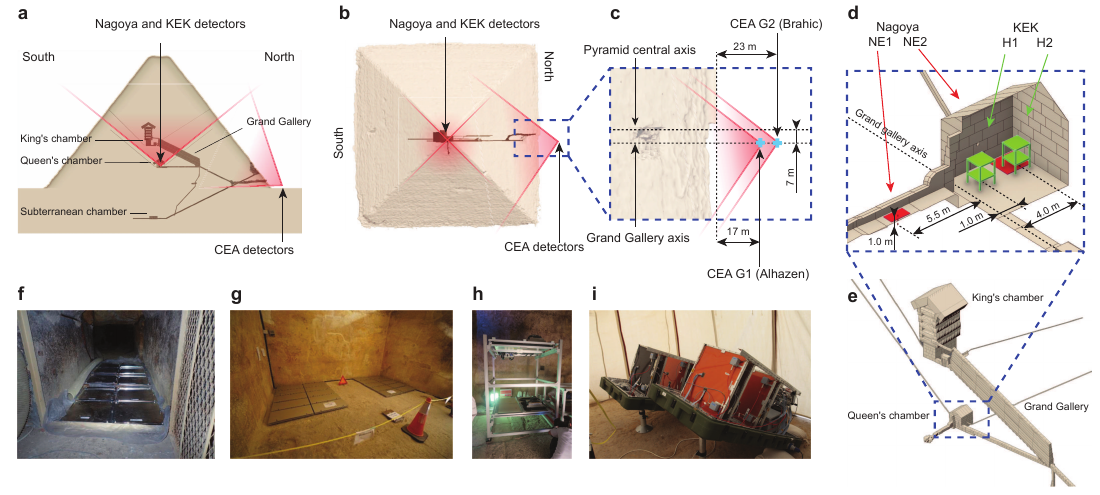}
    \caption{\textbf{Muon detectors installed for Khufu's Pyramid}.\textbf{ 
        a}, Side view of the pyramid, with sensor positions and indicative field of view. 
        \textbf{b}, Top view. \textbf{c}, Close view of the position of the gas detectors 
        (CEA). \textbf{d}, Orthographic view of Queen's chamber with nuclear emulsion films 
        (Nagoya University, red positions NE1 and NE2) and scintillator hodoscopes (KEK, 
        green positions H1 and H2). \textbf{e, }Orthographic view of the main known internal 
        structures \textbf{f}, Nuclear emulsion plates in position NE1 (Nagoya University). 
        \textbf{g}, Nuclear emulsion plates in position NE2 (Nagoya University). \textbf{h}, 
        Scintillator hodoscopes setup for position H1 (KEK). \textbf{i}, Gas detectors 
        (muon telescopes, CEA).}
\end{figure*}

Most of the current understanding of Khufu's Pyramid comes from architectural surveys 
and comparative studies with other pyramids\textsuperscript{1,2,13}. In \textit{Histories}, 
Herodotus described the construction of Khufu's Pyramid, but this account was written 
about 2,000 years later (in 440 BC). The only known documents written during Khufu's 
reign were discovered in 2013\textsuperscript{16}, but these papyri describe only 
the logistics of the construction, such as how the stones were transported, and 
not the construction itself. In 1986, a team surveyed the pyramid using microgravimetry\textsuperscript{17}, 
that is, the measurement of slight variations in gravity due to large variations 
in the amount of matter\textsuperscript{18}. Based on these data, the team drilled 
3 holes in the corridor to the Queen's chamber in the hope of finding a ``hidden 
chamber''; the team only observed sand\textsuperscript{17}. A more recent analysis 
of the same data dismissed the theory of a ``hidden chamber'' where the holes had 
been drilled\textsuperscript{17}.In 1988, a Ground-Penetrating 
Radar survey\textsuperscript{19} suggested that an unknown corridor could be parallel 
to the Queen's chamber corridor. To our knowledge, this theory has been neither 
confirmed nor refuted.

Here we follow in the footsteps of Alvarez \textit{et al.}\textsuperscript{5} 
who used spark chambers as muon detectors in Khafre's pyramid (Kephren) and concluded 
that there is no unknown structure with a volume similar to the 
King's chamber above the Belzoni chamber\textsuperscript{5}. Muon particles originate 
from the interactions of cosmic rays with the atoms of the upper atmosphere, and 
they continuously reach the Earth with a speed near to that of light and a flux 
of around 10,000 per m\textsuperscript{2}per minute\textsuperscript{4}. Similar 
to X-rays which can penetrate the body and allow bone imaging, these elementary 
particles can keep a quasi-linear trajectory while going through hundreds of meters 
of stone before decaying or being absorbed. By recording the position and the direction 
of each muon that traverses their detection surface, muon detectors can distinguish 
cavities from stones: while muons cross cavities with practically no interactions, 
they can be absorbed or deflected by stones. Put differently,{\small{} }muons traversing 
a region with lower-than-expected density will result in a higher-than-expected 
flux in the direction of the region. In the recent years, muon detectors have been 
successfully deployed in particle accelerators, in volcanology\textsuperscript{6}, 
to visualize the inner structure of the Fukushima's nuclear reactor\textsuperscript{11,20}, 
and for homeland security\textsuperscript{21}. In heritage buildings, detectors 
have been recently set up in archaeological sites near Rome\textsuperscript{22} 
and Naples\textsuperscript{23} (Italy), where they were able to detect some known 
structures from underground, and in the Teotihuacan Pyramid of the Sun (Mexico)\textsuperscript{24}. 
However, since the muons generated by cosmic rays come from the sky, the detectors 
can only detect density variations in some solid angle above them (the exact acceptance 
depends on the detection technology). In addition, the muon imaging technique measures~the 
average density of the structure in a given direction: while a void results in 
a lower density, its impact on the muon radiography is determined by the ratio 
between the void and the length of material that is traversed. For this reason, 
small cavities like air shafts cannot be detected by this technique within a reasonable 
exposure time. Lastly, in large and dense buildings like Khufu's Pyramid where 
only about 1\% of muons reach the detectors, the data need to be accumulated over 
several months.

We first installed nuclear emulsion films (Fig. 1f-g), developed by Nagoya University, 
because they are compact and do not require electric power, which makes them well 
suited for installation in the pyramid (Extended Data Table 1). A nuclear emulsion 
film is a special photographic film that can detect muon trajectories in three-dimensional 
images with sub-micrometric accuracy\textsuperscript{7,8,9}. In these experiments, 
we used unit films of 30 × 25 cm\textsuperscript{2}and covered a maximum 8 m\textsuperscript{2} 
at a time. Each film has a 70 microns thick emulsion coating on both sides of 
a 175 microns thick transparent plastic base (Methods). The muon measurement accuracy 
is around 1 micron in position and around 1.8 mrad (\textasciitilde{}0.1degree) 
for vertical track by using the information obtained with one film (Extended Data 
Table 1). 

\begin{figure*}[ht!]
    \begin{center}
        \includegraphics[angle=90]{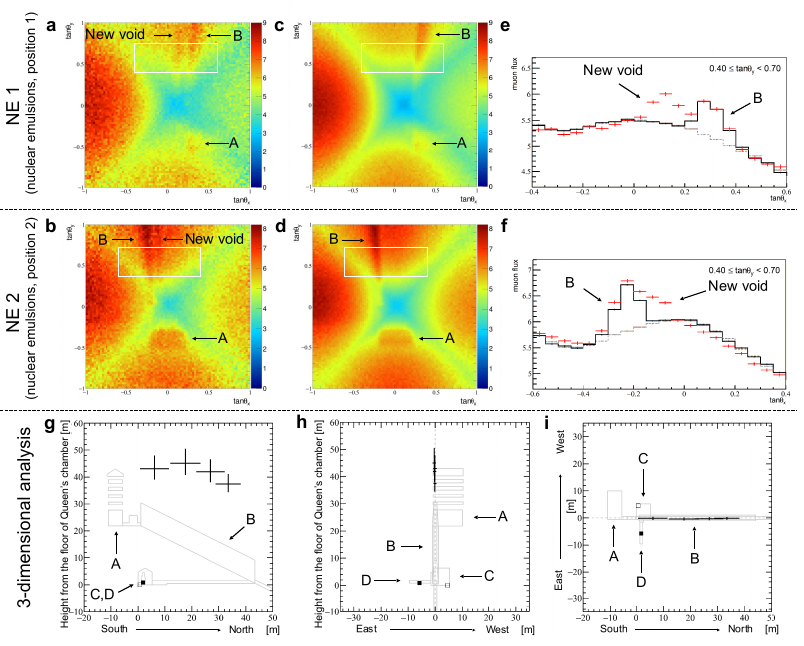}
    \end{center}
    \caption{\textbf{Results of the analysis 
    of the nuclear emulsion films}. (A: King's chamber, B: Grand Gallery, C: Queen's 
    chamber (position NE2), D: Niche (position NE1), New Void: the unexpected muon 
    excess region; these are common in all plots) \textbf{a,b,} Two-dimensional histograms 
    of the detected muon flux (/cm\textsuperscript{2}/day/sr) at positions NE1 and 
    NE2. The resolution of this histogram is 0.025 × 0.025 (Methods). In this figure, 
    right is west, top is north. The four edges of the pyramid are clearly seen as 
    a cross pattern. \textbf{c,d}, Results of simulation with the known inner structures 
    using the Geant4 software\textsuperscript{28} from positions NE1 and NE2. \textbf{e,f,} 
    Histogram of typical angular region as shown by the white square (0.4 $\leq$ tan $\theta$\textsubscript{y} 
    \texttt{<} 0.7). The red line shows the data; the black solid line shows the simulation 
    with the inner structures; the gray dashed line shows the simulation without the 
    inner structures. Error bars indicate 1 sigma of statistical error (standard deviation). 
    More slices are available on the Extended Data Fig. 2. \textbf{g-i,} Results of 
    the triangulation analysis (three sectional views). Each figure shows the inner 
    structures (gray line) and the results. For each position, we divided the region 
    of interest (0 $\leq$ tan $\theta$\textsubscript{y} \texttt{<} 1) into four slices 
    and extracted the center of the muon excess for each of them, resulting in 4 pairs 
    of direction (Methods). Each of the four points represents the result of the triangulation 
    for a pair of slices and the associated statistical error (Methods). The detector 
    positions are shown as a black box for position NE1 and a white box for position 
    NE2; \textbf{g} shows a vertical section (right is north); \textbf{h} shows a vertical 
    section (right is west); \textbf{i }shows a horizontal section (up is west, right 
    is north).}
\end{figure*}

The films were installed near the south-west corner of the Queen's chamber (position 
NE2, Fig.~1d) and in the adjacent narrow hand-excavated corridor called ``Niche'' 
(position NE1, Fig 1d) on the east side of the Queen's chamber. The distance between 
the centres of these two detectors is about 10 m on average, which allows us to 
perform a stereo analysis of the detected structures. The exposure started in December 
2015. During the exposure, we modified the film configuration several times when 
we changed the films (every two months on average); only a subset of the full dataset 
was used for this analysis to mitigate the effect of the parallax on the resolution 
(Extended Data Table 1 and Methods). After each exposure, the nuclear emulsion 
films were processed by photographic development in the dark room of the Grand 
Egyptian Museum Conservation Center. After the development, they were transported 
to Nagoya University and read-out by an automated nuclear emulsion scanning system\textsuperscript{25,26,27}.

We compared the resulting muon radiographs (Fig. 2a,b) with the expected results 
computed using Monte Carlo simulation that include known structures inside the 
observation region (Fig. 2c,d). These comparisons clearly show that the large known 
structures (the Grand Gallery and the King's chamber) are observed by the measurements 
and the results match the expectation (Fig. 2a-d). However, from the two positions, 
we also detected an unexpected and significant excess of muons in a region almost 
parallel to the image of the Grand Gallery. Statistical significance of the excess 
is higher than 10 sigma at the highest difference direction (Fig. 2c). The muon 
excess is similar to the one generated by the Grand Gallery, which means that the 
volume of the two voids is of the same order (Fig. 2e,f and Extended Data Fig. 
2). We then performed a triangulation using the data from the two positions and 
four points along the new void (Methods). The results show that the new void has an estimated length of more than 30 m and is located between 
40 m and 50 m away from the detector positions, 21 m above the ground level.

\begin{figure*}
    \begin{center}
        \includegraphics{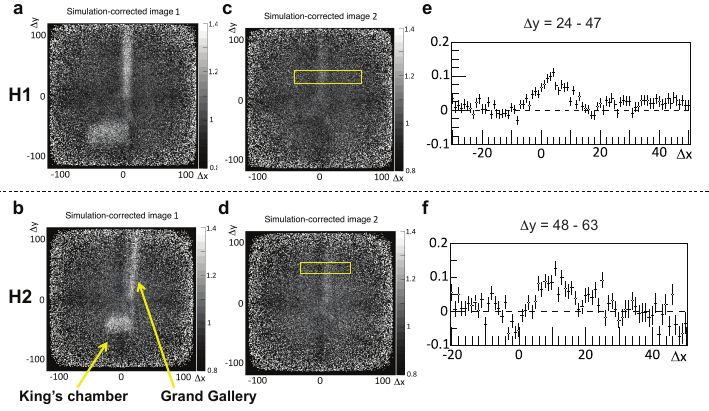}
    \end{center}
    \caption{\textbf{Results of the analysis of the scintillation hodoscopes.} 
    \textbf{a-b,} Two-dimensional histograms of the detected angle of muons after normalization 
    by the simulation without inner structures at positions H1 and H2 (Methods). The 
    images present a top view and the north is the upward. $\Delta$x and $\Delta$y 
    correspond to the channel number difference between the upper and lower layers 
    along x and y coordinates (Methods). The bin number hence ranges from -120 to 120 
    and provides the tangent of the incident angle when divided by 150 (resp. 100) 
    for position H1 (resp. H2).\textbf{ }The detector introduces a dark cross-shaped 
    artefact visible on the two-dimensional histograms, which adds a small systematic 
    error of 3\% to the analysis (Methods)\textbf{. c}-\textbf{d}, Two-dimensional 
    histograms of detected angle of muons after a normalization by the simulation with 
    inner structures (which are therefore removed). The images present the top view 
    and the north is the upward. The unexpected structure is visible. \textbf{e}-\textbf{f,} 
    Histograms of typical angular region (-25 \texttt{<} $\Delta$x \texttt{<} 75, 25 
    \texttt{<} $\Delta$y \texttt{<} 47) corresponding to a slice of panels \textbf{c} 
    and \textbf{d} (yellow rectangle). More slices are available in Extended Data Fig. 
    4. Black points with error bars show the relative excess after dividing the data 
    by the model with the known structures (the King's chamber and the Grand Gallery): 
    a perfect agreement between the data and the model would correspond to a horizontal 
    line; a peak corresponds to an unexpected excess of muons in the data.}
\end{figure*}
The second detection technology, designed by KEK, is composed of four layers of 
scintillator hodoscopes arranged in two units of double orthogonal layers\textsuperscript{10,11} 
(Fig. 1h). Each layer is composed of 120 plastic scintillator bars measuring 1×1 
cm\textsuperscript{2} in cross section to cover an area of 120 × 120 cm\textsuperscript{2}. 
Two units are vertically separated to allow the measurement of the two-dimensional 
incoming direction of the muons. We placed the detectors near the South-East corner 
(position H1, Fig 1d) of the Queen's chamber in August 2016. The separation between 
two units was set to 1.5 m. Unfortunately, the newly detected void was overlapping 
with the Grand Gallery, which made it difficult to identify. After a stable operation 
for five months, we moved the detector to another position near the south-west 
corner of the chamber in January 2017 (position H2, 2.9 m from position H1, Fig. 
1d) and reduced the unit separation to 1.0 m to enlarge the angular coverage of 
the measurement. The detector operation has been very stable for more than one 
year and is still being continued (Methods).

When we normalize the results by the simulation of the solid pyramid without the 
known structures, we see them clearly (Fig. 3a,b). By normalizing with a simulation 
that includes the known structures, we observe a muon excess that is consistent 
with Nagoya's result (Methods, Fig. 3c-f).

The third kind of instrument, designed by the CEA, is made of micro-pattern gaseous 
detectors (Micromegas) based on an argon mixture\textsuperscript{12} (Methods). 
They are robust enough to be installed outside and can run for unlimited time, 
but they have a larger footprint than the emulsion plates (Fig. 1i, Extended Data 
Table 1). Each ``telescope'' is built from four identical detectors with an active 
area of 50 × 50 cm\textsuperscript{2}, and a signal on at least three of them 
is required to trigger the acquisition. An online analysis is performed to extract 
the muon track parameters which are then transferred at CEA, in France, with a 
3G connection (Methods).

\begin{figure*}[ht!]
    \includegraphics{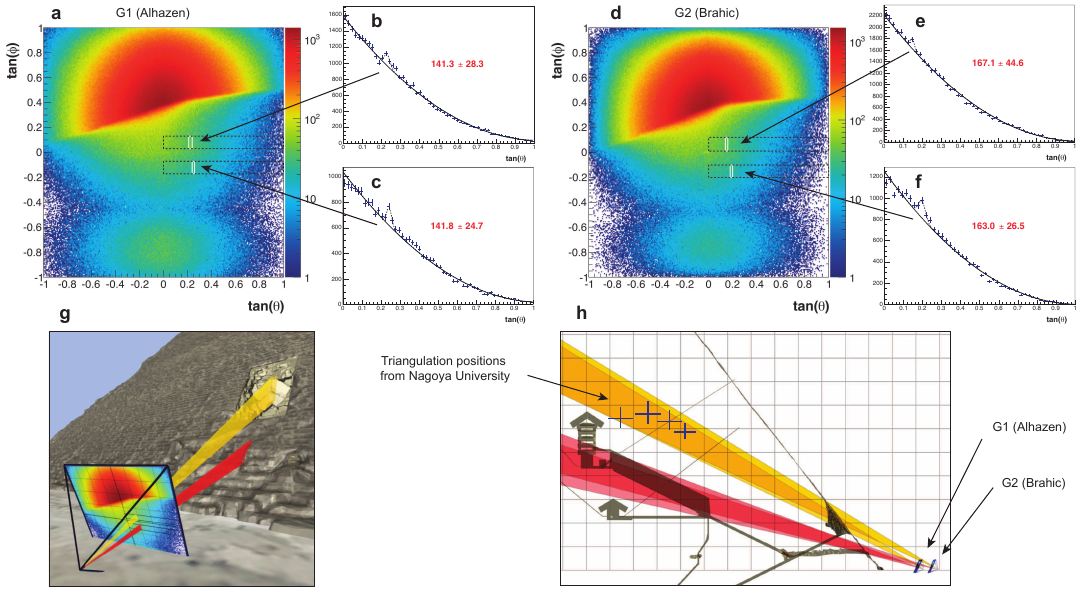}
    \caption{\textbf{Results of the analysis of the gas detectors. 
    a, }2D image from position G1 (Alhazen telescope, Fig. 1a,b) in logarithmic scale,~with 
    white rectangles indicating the 2 muon excesses seen by the telescope. \textbf{b}, 
    Horizontal slice showing the~muon~excess that corresponds~to the~upper dotted black 
    rectangle in panel \textbf{a}. The solid black line is a fit of the data (Methods). 
    \textbf{c}, Horizontal slice showing~the muon~excess~that corresponds to the lower 
    dotted black rectangle in (a). \textbf{d}, 2D image from position G2 (Brahic telescope, 
    Fig 1a,b) in logarithmic scale,~with white rectangles indicating the 2 muon excesses 
    seen by the~telescope. Some noise is visible at the centre of the image, but it 
    does not interfere with the scanned regions. \textbf{e}, Horizontal slice showing 
    the~muon~excess that corresponds to the~upper black rectangle in (d). \textbf{f}, 
    Horizontal slice showing the~muon~excess that corresponds to the~lower black rectangle 
    in (d). \textbf{g}, Yellow cone and red cone, traced from position 1, indicating 
    the angular area for the~observed muon excesses (a-c). The two cones intersect 
    the vertical plane of the Grand Gallery (RTMS software). \textbf{h}, Side view 
    of the same cones as in g, for the two positions,~confirming that the 2 excesses 
    correspond to the Grand Gallery and the new void. The black crosses correspond 
    to the position estimated with the nuclear emulsion plates (Fig. 2g-i): the two 
    analyses are perfectly compatible. Grid size is 10 m (RTMS software). More slices 
    are available on the Extended Data Fig. 6.}
\end{figure*}
In order to confirm Nagoya's discovery and to provide an additional point of view, 
we placed two such telescopes in front of the North face of the pyramid (Fig. 1a-c), 
looking in the direction of the Grand Gallery (Methods), and close enough to each 
other so that their data can be combined (Fig. 1c). A previous 3-month measurement 
campaign with one telescope on the north side already revealed an anomaly in this 
region, above 3 sigma, but the telescope was shifted from the main vertical axis, 
and so not optimally positioned. After two months of acquisition from the new position, 
the data analysis reveals two statistically significant excesses of muons in the 
core of the pyramid (Fig. 4): one corresponding to the Grand Gallery and one corresponding 
to the unexpected void (Methods). The 3D model confirms that the telescopes' observation 
converges to the same region as the one obtained from the Queen's chamber with 
emulsion plates (Fig. 4h). The overall combined excesses yield 8.4 (resp. 5.8) 
sigma for the Grand Gallery (resp. the new void) (Methods). To our knowledge, this 
is the first time an instrument has detected a deep void from outside a pyramid.

Three techniques of cosmic-ray muon imaging were applied to investigate the inner 
structure of the pyramid. The known voids (King's chamber and Grand Gallery) are 
observed as well as an unexpected big void, which fully demonstrates the ability 
of cosmic-ray muon radiography to image structures. The centre of the void is located 
between 40 m and 50 m from the floor of the Queen's chamber. Its length is more 
than 30 m and its cross section is comparable to that of the Grand Gallery. There 
are still many architectural hypotheses to consider; in particular, the big void 
could be made of one or several adjacent structures, and it could be inclined or 
horizontal. The detailed structure of the void should be further studied. Overall, 
this discovery shows how the methods developed in particle physics can shed light 
on one of the most important heritage buildings, and it calls for more interdisciplinary 
collaborations to help understanding the pyramid and its construction process.

\bigskip

\begin{small}
    \sffamily
\begin{itemize}
\item[1.] Lehner, M., \textit{The complete pyramids: Solving the Ancient Mysteries} (Thames 
and Hudson, 2008).

\item[2.] Hawass, Z., Pyramid Construction: New Evidence Discovered in Giza. \textit{Stationen: 
Beiträge zur Kulturgeschichte}, 1998.

\item[3.] Smith, C. B., Hawass, Z. \& Lehner, M., \textit{How the great pyramid was built} 
(Harper Collins, 2006).

\item[4.] Particle Data Group, Review of particle physics. \textit{Chinese Physics} \textbf{C40} 
(10), 100001 (2016).

\item[5.] Alvarez, L. W. \textit{et al.}, Search for hidden chambers in the pyramids. 
\textit{Science} \textbf{167} (3919), 832-839 (1970).

\item[6.] Tanaka, H. K. M., Nakano, T., Takahashi, S., Yoshida, J. \& Niwa, K., Development 
of an emulsion imaging system for cosmic-ray muon radiography to explore the internal 
structure of a volcano, , Mt. Asama. \textit{Nucl. Instrum. Methods Phys. Res.} 
\textbf{A575}, 489-497 (2007).

\item[7.] Morishima, K., Nishio, A., Moto, M., Nakano, T. \& Nakamura, M., Development 
of nuclear emulsion for muography. \textit{Annals of Geophysics} \textbf{60} (1), 
0112 (2017).

\item[8.] Nakamura, T. A. \emph{et al.}, The OPERA film: New nuclear emulsion for large-scale, 
high-precision experiments.\texttt{"}~. \textit{Nuclear Instruments and Methods 
in Physics Research Section A: Accelerators, Spectrometers, Detectors and Associated 
Equipment~} \textbf{556} (1), 80-86 (2006).

\item[9.] Nishio, A., Morishima, K., Kuwabara, K. \& Nakamura, M., Development of nuclear 
emulsion detector for muon radiography. \textit{Physics Procedia} \textbf{80}, 
74-77 (2015).

\item[10.] Fujii, H. \emph{et al.}, Detection of on-surface objects with an underground radiography 
detector system using cosmic-ray muons. \textit{PTEP} (5), 053C01 (2017).

\item[11.] Fujii, H. \emph{et al.}, Performance of a remotely located muon radiography system 
to identify the inner structure of a nuclear plant. \textit{PTEP} \textbf{7}, 073C01 
(2013).

\item[12.] Bouteille, S. \textit{et al.}, A Micromegas-based telescope for muon tomography: 
The WatTo experiment (and references therein). \textit{Nuclear Instruments and 
Methods in Physics Research Section A: Accelerators, Spectrometers, Detectors and 
Associated Equipment} \textbf{834}, 223-228 (2016).

\item[13.] Dash, G., The Great Pyramid's Footprint: Results from our 2015 Survey. \textit{Aeragram} 
\textbf{16} (2), 8-14 (2015).

\item[14.] Hawass, Z. \textit{et al.}, First report: Video survey of the southern shaft 
of the queen's chamber in the great pyramid. \textit{Annales du Service des Antiquités 
de l'Egypte} \textbf{84}, 203-216 (2010).

\item[15.] Richardson, R. \textit{et al.}, The \texttt{"}Djedi\texttt{"} Robot Exploration 
of the Southern Shaft of the Queen's Chamber in the Great Pyramid of Giza, Egypt. 
\textit{Journal of Fields Robotics} \textbf{30} (3), 323-348 (2013).

\item[16.] Tallet, P., \textit{Les papyrus de la Mer Rouge 1: Le \texttt{"}journal de 
Merer\texttt{"}} (Institut Français d'Archéologie Orientale, 2017).

\item[17.] Bui, H. D., \textit{Imaging the Cheops pyramid} (Springer Science \& Business 
Media, 2011).

\item[18.] Butler, D. K., Microgravimetric and gravity gradient techniques for detection 
of subsurface cavities. \textit{Geophysics} \textbf{49} (7), 1084-1096 (1984).

\item[19.] Yoshimura, S., Nakagawa, T., Tonouchi, S. \& Seki, K., Non destructive Pyramids 
investigation, Part 1 and 2. \textit{Studies in Egyptian Culture} \textbf{6} (1987).

\item[20.] Morishima, K. \textit{et al.}, First demonstration of cosmic ray muon radiography 
of reactor cores with nuclear emulsion based on an automated high-speed scanning 
technology. \textit{Proc. of the 26th Workshop on Radiation Detectors and Their 
Uses} (2012).

\item[21.] Borozdin, K. N. \textit{et al.}, Surveillance: Radiographic imaging with cosmic-ray 
muons. \textit{Nature} \textbf{422} (6929), 277 (2003).

\item[22.] Menichelli, M. \textit{et al.}, A scintillating fibres tracker detector for 
archaeological applications. \textit{Nuclear Instruments and Methods in Physics 
Research Section A: Accelerators, Spectrometers, Detectors and Associated Equipment} 
\textbf{572} (1), 262--265 (2007).

\item[23.] Saracino, G. \textit{et al.}, Imaging of underground cavities with cosmic-ray 
muons from observations at Mt. Echia (Naples). \textit{Scientific Reports} \textbf{7} 
(1), 1181 (2017).

\item[24.] Menchaca-Rocha, A., Searching for cavities in the Teotihuacan Pyramid of the 
Sun using cosmic muons experiments and instrumentation. \textit{International Cosmic 
Ray Conference} \textbf{4}, 325 (2011).

\item[25.] Yoshimoto, M., Nakano, T., Komatani, R. \& Kawahara, H., Nuclear emulsion readout 
system HTS aiming at scanning an area of one thousand square meters. \textit{e-Print: 
arXiv}, 1704.06814 (2017).

\item[26.] Morishima, K., Hamada, K., Komatani, R., Nakano, T. \& Kodama, K., Development 
of an automated nuclear emulsion analyzing system. \textit{Radiation Measurements} 
\textbf{50}, 237-240 (2013).

\item[27.] Hamada, K. \textit{et al.}, Comprehensive track reconstruction tool \texttt{"}NETSCAN 
2.0\texttt{"} for the analysis of the OPERA Emulsion Cloud Chamber. \textit{Journal 
of Instrumentation} \textbf{7} (7), P07001 (2012).

\item[28.] Agostinelli, S. \textit{et al.}, GEANT4---a simulation toolkit. \textit{Nuclear 
instruments and methods in physics research section A: Accelerators, Spectrometers, 
Detectors and Associated Equipment} \textbf{506} (3), 250-303 (2003).
\end{itemize}
\end{small}
\bigskip

\small
\sffamily

\noindent\textbf{Author contributions} K. M., M. K., A. N., N. K., Y. M. and M. M. performed 
the experiments and analysed the results for the nuclear emulsion films; F. T., 
H. F., K. S., H. K., K. H. and S. O. performed the experiments and analysed the 
results for the scintillator hodoscopes. S. P., D. A., S. B., D. C., C. F., P. 
M., I. M. and M. R. performed the experiment and analysed the results for telescopes 
gas detectors. B. M., P. G., E. G., N. S., Y. D. and M. S. created the 3D models 
used for muography simulations and RTMS. B. M. designed and implemented the Real 
Time Muography Simulator (RTMS) and contributed to the analyses; Y. E., T. E., 
M. E. and V. S. coordinated the different experiments operations on the field (muography, 
3D scans); the paper was mainly written by K. M., S. P., F. T., M. T., B. M. and 
J.-B. M., with contributions of all the other authors; H. H., M. T., B. C., B. 
M. and Y. E. designed and coordinated the project (ScanPyramids);

\smallskip

\noindent\textbf{Author information }
Correspondence and requests for materials should be addressed both to K. M. (morishima@flab.phys.nagoya-u.ac.jp) 
and M. T. (tayoubi@hip.institute).

\smallskip

\noindent\textbf{Acknowledgements} This experiment is part of the {ScanPyramids} project, 
which is supported by NHK, La Fondation Dassault Systèmes, Suez, IceWatch, le 
Groupe Dassault, Batscop, Itekube, Parrot, ILP, Kurtzdev, Gen-G, Schneider Electric.~The 
measurement with nuclear emulsions was supported by the JST-SENTAN Program from 
Japan Science and Technology Agency, JST and JSPS KAKENHI Grant Numbers JP15H04241.~The 
CEA telescopes were partly funded by the Région Ile-de-France and the P2IO LabEx 
(ANR-10-LABX-0038) in the framework ``Investissements d'Avenir'' (ANR-11-IDEX-0003-01) 
managed by the Agence Nationale de la Recherche (ANR, France). The~detectors were 
built~by the ELVIA company and the CERN MPGD workshop.~ The authors would like 
to thank the members and benefactors of the ScanPyramids project, and in particular:~T. 
Hisaizumi,~the members of Cairo University, the members of F-lab Nagoya University, 
Y. Doki, Aïn El Shams University 3D scanning team, the members of Egyptian Ministry 
of Antiquities, K. El Enany, M. El Damaty, T. Tawfik, S. Mourad, S. Tageldin, E. 
Badawy, M. Moussa, T. Yabuki, D. Takama, T. Shibasaki, K. Tsutsumida, K. Mikami, 
J. Nakao, H. Kurihara, S. Wada, H. Anwar, T. de Tersant, P. Forestier, L. Barthès, 
M.-P. Aulas, P. Daloz, S. Moignet, V. Raoult-Desprez, S. Sellam, P. Johnson, J.-M. 
Boursier, , T. Alexandre, V. Ferret, T. Collet, H. Andorre, C. Oger-Chevalier, 
V. Picou, B. Duplat, K. Guilbert, J. Ulrich, D. Ulrich, C. Thouvenin, L. Jamet, 
A. Kiner, M.-H. Habert, B. Habert, L. Gaudé, F. Schuiten, F. Barati, P. Bourseiller, 
R. Theet, J.-P. Lutgen, R. Chok, N. Duteil, F. Tran, J.-P. Houdin, L. Kaltenbach, 
M. Léveillé-Nizerolle, R. Breitner, R. Fontaine, H. Pomeranc, F. Ruffier, G. 
Bourge, R. Pantanacce, M. Jany, L. Walker, L. Chapus, E. Galal, H. A. Mohalhal, 
S. M. Elhindawi, J. Lefaucheux, J.-M. Conan, E. M. Elwilly, A. Y. Saad, H. Barrada, 
E. Priou, S.Parrault, J.-C. Barré, X. Maldague, C. Ibarra Castenado, M. Klein, 
F. Khodayar, G. Amsellem, M. Sassen, C. Béhar, M. Ezzeldin, E. Van Laere, D. Leglu, 
B. Biard, N. Godin, P. der Manuelian, L. Gabriel, P. Attar, A. De Sousa, F. Morfoisse, 
R. Cotentin.

\bigskip

\section*{Methods}

\noindent\textbf{Detector Design (nuclear emulsions - Nagoya University)}. Nuclear emulsion 
is a special photographic film that is able to detect minimum ionizing particles 
such as cosmic-ray muons (Extended Data Fig. 1). The films used in this experiment 
have been developed and produced at Nagoya University. In this design, silver bromide 
crystals with a diameter of 200 nm are dispersed in a 70 $\mu$m thick sensitive 
emulsion layer, which is coated on both sides of a 175 $\mu$m thick transparent 
polystyrene plastic base\textsuperscript{7,29} (Extended Data Fig. 1a,b). When 
a charged particle passes through this emulsion layer, its three-dimensional trajectory 
(track) is recorded and can be revealed through the chemical development process 
(Extended Data Fig. 1c,e). Thanks to the precise grain size and structure, tracks 
can be reconstructed with sub-micron accuracy in 4$ $$\pi$ steradians by using 
an optical microscope (Extended Data Fig. 1d). The track reconstruction quality 
depends on the grain density per length along the line of a track. In this experiment, 
films with a grain density of 37 grains per 100 $\mu$m and a noise level of about 
1 grain per 1000 $\mu$m\textsuperscript{3} were used. These films can be used for 
long term (2-3 months) measurement in an environment at 25 °C (temperature in 
the Queen's chamber) by tuning of volume occupancy of silver bromide crystals (35\%)\textsuperscript{30}. 
Each 30×25 cm\textsuperscript{2} film was vacuum-packed in an aluminium laminated 
package for light shielding and humidity control (30\%RH) (Extended Data Fig. 1f). 
An acrylic plate with 2 mm thickness was also packed together for the control of 
noise increase\textsuperscript{9}. The packed film was fixed onto an aluminum honeycomb 
plate (Extended Data Fig. 1 g,h) at Nagoya University, and then transported to 
Cairo by airplane. To avoid the elevated cosmic-ray flux during the flight, we 
transported the two emulsions layers of each detector separately and assembled 
them in Egypt. Since we reject the trajectories that are not crossing the two layers, 
the muons accumulated during the flight are rejected in the analysis and are considered 
as background tracks.

\smallskip

\noindent\textbf{Data acquisition (Nuclear emulsions - Nagoya University).} The observation 
with nuclear emulsion films was launched in the Queen's chamber in December 2015 
and the films were periodically replaced. Each film was aligned with a reference 
line drawn with a laser marker and a spirit level, which led to an angular error 
of less than 10 mrad. After each exposure, we processed the films by photographic 
development in the dark room at the Grand Egyptian Museum Conservation Center. 
For the developing solution, we used the XAA developer (FUJI FILM Co. Ltd.) for 
25 minutes at 20°C. After the development, we carried the films to Nagoya University, 
where they were read-out by an automated nuclear emulsion scanning system developed 
since early 1980's in Nagoya University\textsuperscript{31,32}. In this experiment, 
tracks recorded in films were scanned by a Hyper Track Selector\textsuperscript{25} 
which can read-out tracks at a speed of 4700 cm\textsuperscript{2}/h with angular 
accuracy of 1.8 mrad for vertical tracks, and saved in a computer storage device 
as digital data (position, angle, pulse height). The angular acceptance is approximately 
\textbar{}tan$\theta$\textbar{} $\leq$ 1.3 where $\theta$ is the zenith angle relative 
to the perpendicular of the emulsion surface.

\smallskip

\noindent\textbf{Data analysis and statistics (nuclear emulsions - Nagoya University)}. 
The muon tracks are reconstructed by the coincidence between the two stacked films 
within criteria of signal selection and then counted as detected muons\textsuperscript{26}. 
In this analysis, a subset of the full data set was used to avoid decreasing the 
resolution because of imaging parallax: 4.4 million tracks were accumulated for 
98 days at position NE1 and 6.2 million tracks for 140 days at position NE2, with 
an effective area of 0.45 m\textsuperscript{2}. Subsequently, detected muons were 
integrated into two-dimensional angular space (tan$\theta$\textsubscript{x}-tan$\theta$\textsubscript{y}) 
with the bin size of a specified size (e.g., 0.025 × 0.025) and the angular acceptance 
of \textbar{}tan$\theta$\textbar{}$\leq$ 1.0, and converted to muon flux (/cm\textsuperscript{2}/day/sr) 
in each bin (Fig. 2).

We used the Monte Carlo simulator Geant4 Version 10.2\textsuperscript{28} to compute 
the expected muon flux at detector position. In these simulations, the physical 
process of electromagnetic interactions and decays of muons were included, Miyake's 
formula of integrated intensity of cosmic-ray muons was utilized as a flux model, 
and only muons were generated as primary particles in this simulator\textsuperscript{33}. 
In order to reduce the processing time, only muons were propagated and the range 
of incident muon energy was limited to be 20 to 1000 GeV in the zenith angular 
range 0 to 70 degree (-2.75 \texttt{<} tan$\theta$ \texttt{<} 2.75).  For the pyramid 
simulation, we modeled the shape and the location of the known structures (the 
Grand Gallery, the King's chamber, the corridor that connects them, and the Queen's 
chamber) using the survey of Dormion\textsuperscript{34}. We defined that any empty 
void would be filled with air and that the stones are limestone (2.2 g/cm\textsuperscript{3}) 
except around the King's chamber, where they are granite (2.75 g/cm\textsuperscript{3}). 
The exposure period in the simulation is compatible with 1000 days, which is approximately 
10 times longer than that of analyzed data. We estimated the rock thickness distribution 
in 2D from the detector position: the minimum thickness is 65 m and the maximum 
thickness is 115 m. If we assume the fluctuation of surface structure is 1 m scale 
(stone size), the effect of the relative fluctuation is less than 2\%.

Normalization was performed to compare real and simulated data in the region without 
the analyzing target (the Grand Gallery, the King's chamber, and the anomaly region). 
The excess region of muon flux was clearly apparent in the images (Figure 2a-d). 
Two histograms (Figure 2e,f) show muon flux extracting from the slice in 0.4 $\leq$ 
tan$\theta$\textsubscript{y} \texttt{<} 0.7. From the comparison between data and 
the simulation, the significances of each anomaly region were evaluated at 13.7 
sigma (statistical) for position NE1 and 12.7 sigma for position NE2.

In order to locate the newly discovered structure, we performed a triangulation 
from the two positions. The centre of detector position NE1 was located at 5.8 
m east from the axis of the Grand Gallery and at 4.5 m west for position NE2. The 
distance between position NE1 and NE2 is 1.1 m in north-south (Figure 1d). In order 
to determine the direction toward the anomaly region, we performed the fitting 
of the excess region to a Gaussian function by dividing the region (0 $\leq$ tan$\theta$\textsubscript{y} 
\texttt{<} 1) into four regions with a segment of 0.25 in tan$\theta$\textsubscript{y}, 
because the new structure seems to align along the tan$\theta$\textsubscript{y} 
axis direction (Extended Data Fig. 2). The fitted center value was used for triangulation 
and the errors of the estimated positions were defined from the errors on the sight 
lines coming from a half of the bin width, i.e. 0.0125 in tan$\theta$\textsubscript{x} 
and 0.125 in tan$\theta$\textsubscript{y} (Fig. 2g-i).

\smallskip

\noindent\textbf{Detector design (scintillator hodoscopes - KEK).} The detector consists 
of two units of double layers, \textit{i.e.}, x and y layers, of plastic scintillator 
arrays (Extended Data Fig. 3a-c). A single scintillator element is 10 ×10 mm\textsuperscript{2} 
in cross section and 1200 mm long. Each layer has 120 elements tightly packed, 
and hence its active area is 1200 × 1200 mm² (Extended Data Fig. 3d). The element 
has a central hole along its length, through which a wave-shifter optical fiber 
is inserted to efficiently transfer the scintillation light to a Multi-Pixel Photon 
Counter Sensor (MPPC, Hamamatsu). The bias voltage of the MPPC was selected according 
to the temperature of the Queen's Chamber, which is constant regardless of the 
weather outside. Each layer has its own DAQ box, which digitizes the information 
of sensor signals and sends them to a common PC inside the detector frame. The 
total power consumption of the detector system is 300 W. The vertical distance 
between the two units is 1500 mm at position H1 and 1000 mm at position H2, and 
gives an angular resolution around 7 and 10 mrad respectively. The tangent acceptance 
ranges from 0 (vertical) to 0.8 and 1.2 rad respectively.

The detector introduces a dark cross-shaped artefact visible on the two-dimensional 
histograms (Fig. 3a-d), which adds a small systematic error of 3 \% to the analysis\textbf{. 
}According to our analyses, the error is likely to be caused by the very narrow 
gap between neighboring scintillator elements, but this effect has not been fully 
understood yet. This systematic error is not relevant in the present analysis, 
which only examines the existence of the new void.

\smallskip

\noindent\textbf{Data acquisition (scintillator hodoscopes - KEK). }Raw data - time and 
position of all hit channels - are first stored in a PC and regularly retrieved 
with USB memory to be sent to KEK through the Internet.\textbf{ }In the off-line 
analysis, a muon event is defined by the coincidence of the 4 layers, with at most 
2 neighboring hits in each of them. Events are accumulated in two-dimensional bins 
($\Delta$x, $\Delta$y) given by the channel number differences between the upper 
and lower layers along x and y coordinates. The bin number hence ranges from -120 
to 120 and provides the tangent of the incident angle when divided by 150 (resp. 
100) for position H1 (resp. H2). We installed the detector at position H1 in August 
2016, and continued the data taking for five months until January in 2017. During 
this period, we accumulated 4.8 M events. We then moved the detector by 2.9 m west 
in order to better investigate the newly observed void. The data taking is still 
continuing for more than eight months and 12.9 M events were accumulated at position 
H2 at the end of September 2017, with an overall smooth acquisition for more than 
a year.

\smallskip

\noindent\textbf{Data analysis and statistics (scintillator hodoscopes - KEK). }The first 
step of the analysis is the normalization of the data by a Monte Carlo simulation 
that takes into account the cosmic ray muon flux and muon interactions\textsuperscript{35,36 
}(energy loss and multiple scattering) in the Pyramid. We assume a constant energy 
loss of 1.7 MeV per g/cm\textsuperscript{2 (ref. 4)}, a mean density of 2.2 g/cm\textsuperscript{3}, 
and a radiation length of 26.5 g/cm\textsuperscript{2} for the stones. Muons are 
propagated in steps of 0.1 m. Because the known structures of the Pyramid are simulated, 
their effects are removed after the normalization of the data and the remaining 
muon excess shows the existence of an unknown corridor-like new structure. The 
successful elimination of the known structures suggests the reliability of our 
simulation.\textbf{ }Slices of the images along $\Delta$x are presented in Extended 
Data Fig. 4a,b. The vertical scale is the relative yield to the simulation result. 
The new structure is clearly seen in each slice. The significance of the muon excess 
was obtained by a Gaussian fit: at position H1 the excess heights in the slices 
ranges from 5.2\% to 8.9\% and more than 10 sigma except for the most outer slice, 
in which the height is still more than 5 sigma. At position H2 the height ranges 
from 8.9\% to 11\% and are again above 10 sigma except for the most outer slice, 
in which the excess is above 7 sigma. From these slices, we found that the structure 
starts almost at the centre of the Pyramid and ends at an angle whose tangent is 
0.8 to the North. As a result, the length of the main part of the new structure 
is approximately 30 m. Results from both position H1 and H2 show that the new void 
is above the Grand Gallery, which is consistent with Nagoya's result.

\smallskip

\noindent\textbf{Detector design (gas detectors - CEA).} A telescope (Extended Data Fig. 
5a) is composed of 4 Micromegas (Extended Data Fig. 5b), a Micro-Pattern Gaseous 
Detector (MPGD) invented at CEA-Saclay\textsuperscript{37} (Extended Data Fig. 
5c). All the detectors are identical, with an active area of 50 × 50 cm² \textsuperscript{(ref. 
12)}. They are built from the bulk technology\textsuperscript{38} with a screen-printing 
resistive film on top of the readout strips to allow for stable operation and higher 
gain\textsuperscript{39}. Each detector provides X and Y coordinates through a 
2D readout inserted onto the printed circuit board (Extended Data Fig. 5d). The 
1037 readout strips (482 micron pitch) of each coordinate are multiplexed according 
to a patented scheme\textsuperscript{40}. An Argon-iC4H10-CF4, non-flammable gas 
mixture (95-2-3) is flushed in series in all the detectors of a telescope, with 
a flow below 0.5 L/h, thanks to a tight seal (measured gas leak below 5 mL/h per 
detector).

Each telescope is operated with a Hummingboard nano-PC running GNU/Linux, which 
controls all the electronics\textsuperscript{41}: a dedicated high voltage power 
supply (HVPS) card with 5 miniaturized CAEN modules, which provide up to 2 kV with 
12V inputs, and the Front End Unit (FEU) readout electronics based on the DREAM 
ASIC\textsuperscript{42}. A particularly important feature of DREAM is its self-triggering 
option to generate the trigger from the detectors themselves. A dedicated software 
package was developed to interface all these electronic components to the Hummingboard 
which performs the data acquisition with the FEU. It also monitors and sets the 
high voltages through the HVPS and a patented amplitude feedback to keep the gain 
constant in spite of the extreme environmental conditions of the Giza plateau (Extended 
Data Fig. 5 e-g). The overall consumption of a telescope is only 35 W.

A trigger is formed by the FEU when at least 5 coordinates out of 8 observe a signal 
above a programmable threshold. The sampled signals (50 samples at a frequency 
of 100/6 MHz, Extended Data Fig. 5h) of all the electronic channels (64 × 8) are 
then directly converted to a ROOT file\textsuperscript{43}, a format commonly used 
in particle physics. The nano-PC performs the online reconstruction of muon trajectories, 
and the muon track parameters are sent to France with some environmental data (temperature, 
pressure, humidity in the gas) through a 3G connection. 

\smallskip

\noindent\textbf{Data acquisition (gas detectors - CEA).} The data were collected from May 
4\textsuperscript{th} to July 3\textsuperscript{rd}, 2017 with a stable data taking. 
Two telescopes were installed in front of the chevrons (North face), at a distance 
of 17 and 23 m, respectively (Fig. 1b,c). The axis of both telescopes deviated 
slightly from the north-south axis, toward the east, to prevent the Grand Gallery 
from being at the centre of the image, where some correlated noise can show up. 
During the acquisition time the two telescopes (called Alhazen and Brahic, at positions 
G1 and G2, see Fig. 1b,c) recorded 15.0 and 14.5 million triggers, respectively, 
from which 10.6 and 10.4 million of track candidates were identified. After the 
chi² quality cut, 6.9 and 6.0 million good tracks were reconstructed, and form 
the images shown in Figure 4 (a) and (d). 

\smallskip

\noindent\textbf{Data analysis and statistics (gas detectors - CEA).} From the acquired 
tracks, we searched for anomalies by extracting slices in tan(phi), i.e. horizontal 
slices. To get more statistics, the thickness of the slices is larger than the 
binning shown in the 2D images. We chose a slice thickness of 0.10 for the Alhazen 
telescope, and we increased to 0.11 for the Brahic telescope to keep roughly the 
same statistics. The Extended Data Fig. 6 illustrates all the slices made with 
Alhazen from 0.21 to -0.19. From one histogram to another, the slice position is 
shifted by 0.02, which means the data of consecutive histograms largely overlap. 
The goal is to scan the pyramid and detect any deviation from statistics, being 
fluctuations or not.

As can be seen on Extended Data Fig. 6, the slices show smooth distributions, except 
around histograms 5-6 and 15. The histograms 6 and 15 correspond to Fig. 4b and 
4c respectively. These distributions were fitted with different functions, in particular 
polynomials. Though such functions are essentially empirical, a CRY/Geant4 simulation 
was performed to further validate this choice, leading to a good agreement using 
a 2\textsuperscript{nd} order polynomial with a reduced $\chi$² of 1.4.  The same 
function reproduces data distributions fairly well - with a reduced $\chi$² value 
of 1.6 and 2.0 respectively for histograms 6 and 15 - except in a region where 
an excess is clearly observed on both slices, with single bin excess of 4.2 and 
5.3 sigma respectively. Re-fitting with a 2\textsuperscript{nd} order polynomial 
and a Gaussian significantly reduces the $\chi$² to 1.2 and 1.4, with a Gaussian 
integral of 141.3 {\color{color23} ±} 28.3 (5.0 sigma, histogram 6) and 141.8 
{\color{color23} ±} 24.7 (5.7 sigma, histogram 15). Similarly, Brahic data show 
2 significant excesses, corresponding to Fig. 4e and 4f. A 2\textsuperscript{nd} 
order polynomial alone results in reduced $\chi$² value of 1.8 and 2.4 respectively, 
while adding a Gaussian reduces them to 1.5 and 1.6. The Gaussian integral is 167.1 
{\color{color23} ±} 44.6 (3.7 sigma, Fig. 4e) and 163 {\color{color23} ±} 26.5 
(6.1 sigma, Fig. 4f).

The 3D model confirms that the (well compatible) excesses from Fig. 4c (Alhazen) 
and 4f (Brahic) point to the same region of the pyramid and overlap very well with 
the Grand Gallery (Fig. 4h). The quasi-full overlap of the cones (due to the purposeful 
proximity of the telescopes) justifies adding the 2 excesses, leading to 304.8 
{\color{color23} ±} 36.2 (i.e., 8.4 sigma). This fully validates the ability of 
the telescopes to unambiguously detect large structures in the core of the pyramid.

The 3D model also confirms that the excesses from Fig. 4b and 4e point to the same 
region. Like before, the quasi-full overlap of the cones justifies adding the 2 
excesses, leading to 308.4 {\color{color23} ±} 52.8 (5.8 sigma). A 3D comparison 
with the triangulation made by Nagoya University further confirms a large overlap 
of these regions. The other slices show no other anomaly exceeding 5 sigma.

It is worth mentioning that this analysis relies directly on raw data without model 
subtraction, which means the systematics are much smaller, and can only originate 
from the fitting function. As an exercise, a 3\textsuperscript{rd} polynomial fit 
was used for the histograms showing the new void, resulting in excesses of 5.2 
and 3.6 sigma for Alhazen and Brahic, and a combined excess of 303.5 {\color{color23} ±} 
52.1 (5.8 sigma).

\smallskip

\noindent\textbf{3-dimensional model of the pyramid.} We designed an accurate 3-dimensional 
model of the pyramid by combining existing architectural drawings\textsuperscript{34,44 
}photogrammetry\textsuperscript{45}, and laser scanner measurements\textsuperscript{45}, 
both inside and outside of the pyramid. After merging these data, the model contains 
about 500,000 triangles (Extended Data Fig. 7 b-d). This model was mainly used 
in the RTMS simulator (see below) and as reference for the simplified models used 
in the other simulators. The full model has a precision of approximately 30 cm 
for the internal structures and approximately 1 m for the external casing.

\smallskip

\noindent\textbf{Real Time Muography Simulator. }The Real Time Muography Simulator (RTMS, 
Extended Data Figure 7a) is a fast, interactive simulator that was mainly used 
for preliminary analyses, to aid positioning the gas sensors (telescopes), and 
for confirming the results obtained from the others simulators. It allows the user 
to place a sensor in the detailed 3D model of the pyramid and to simulate the observed 
muon rate in real time. Muon scattering is not simulated.\textbf{ }For each pixel 
of the sensor, which represents a direction relative to the sensor, the simulator 
computes the opacity (integral of the density along the path) from the sensor to 
the outside of the pyramid, along this direction\textbf{. }We used a density of 
2.2 g/cm\textsuperscript{3} for the limestone, and 2.6 g/cm\textsuperscript{3} 
for the granite. We consider that muons lose energy at a constant rate of 1.69 
MeV per g/cm\textsuperscript{2} \textsuperscript{(ref 4)} which allows us to compute 
the minimal energy E\textsubscript{min} to cross the pyramid given the value of 
the opacity. Finally, we use Miyake's formula\textsuperscript{33} to calculate 
the distribution of muons that have greater energy than E\textsubscript{min} coming 
at a zenith angle $\theta$. This value is computed for each pixel of the image, 
leading to a 2-dimensional histogram similar to those obtained with the detectors.

\smallskip

\noindent\textbf{Data availability. }The data that support the findings in this study are 
available from the corresponding authors on reasonable request.

\smallskip

\begin{itemize}
\item[29.] Morishima, K., Latest Developments in Nuclear Emulsion Technology. \textit{Physics 
Procedia} \textbf{80}, 19-24 (2015).

\item[30.] Nishio, A. \textit{et al.}, Long Term Property of Nuclear Emulsion. \textit{Program 
and Proceedings of The 1st International Conference on Advanced Imaging (ICAI2015)}, 
668-671 (2015).

\item[31.] Aoki, S. \textit{et al.}, The Fully Automated Emulsion Analysis System. \textit{Nuclear 
Instruments and Methods in Physics Research Section B: Beam Interactions with Materials 
and Atoms} \textbf{51}, 466-472 (1990).

\item[32.] Morishima, K. \& Nakano, T., Development of a new automatic nuclear emulsion 
scanning system, S-UTS, with continuous 3D tomographic image read-out. \textit{Journal 
of Instrumentation} \textbf{5} (4), P04011 (2010).

\item[33.] Miyake, S., Rapporteur paper on muons and neutrinos. \textit{13th International 
cosmic ray conference} (1973).

\item[34.] Dormion, G., \textit{La chambre de Cheops: analyse architecturale} (Fayard, 
2004).

\item[35.] Reyna, D., A simple parameterization of the cosmic-ray muon momentum spectra 
at the surface as a function of zenith angle. \textit{arXiv preprint}, hep-ph/0604145 
(2006).

\item[36.] Jokisch, H., Carstensen, K., Dau, W., Meyer, H. \& Allkofer, O., Cosmic-ray 
muon spectrum up to 1 TeV at 75 zenith angle. \textit{Physical review D} \textbf{19} 
(5), 1368 (1979).

\item[37.] Giomataris, I. Y., Rebourgeard, P., Robert, J. P. \& Charpak, G., MICROMEGAS: 
a high-granularity position-sensitive gaseous detector for high particle-flux environments. 
\textit{Nuclear Instruments and Methods in Physics Research Section A: Accelerators, 
Spectrometers, Detectors and Associated Equipment} \textbf{376} (1), 29-35 (1996).

\item[38.] Giomataris, I. Y. \textit{et al.}, Micromegas in a bulk. \textit{Nuclear Instruments 
and Methods in Physics Research Section A: Accelerators, Spectrometers, Detectors 
and Associated Equipment} \textbf{560} (2), 405-408 (2006).

\item[39.] Alexopoulos, T. \textit{et al.}, A spark-resistant bulk-micromegas chamber 
for high-rate applications. \textit{Nuclear Instruments and Methods in Physics 
Research Section A: Accelerators, Spectrometers, Detectors and Associated Equipment} 
\textbf{640}, 110-118 (2011).

\item[40.] Procureur, S., Dupré, R. \& Aune, S., Genetic multiplexing and first results 
with a 50x50cm$^2$ Micromegas. \textit{Nuclear Instruments and Methods 
in Physics Research Section A: Accelerators, Spectrometers, Detectors and Associated 
Equipment} \textbf{729}, 888-894 (2013).

\item[41.] Bouteille, S. \textit{et al.}, A Micromegas-based telescope for muon tomography: 
The WatTo experiment. \textit{Nuclear Instruments and Methods in Physics Research 
Section A: Accelerators, Spectrometers, Detectors and Associated Equipment} \textbf{834}, 
223-228 (2016).

\item[42.] Flouzat, C. \textit{et al.}, Dream: a 64-channel Front-end Chip with Analogue 
Trigger Latency Buffer for the Micromégas Tracker of the CLAS12 Experiment. \textit{Proc. 
of TWEPP conference} (2014).

\item[43.] Brun, R. \& Rademakers, F., ROOT---an object oriented data analysis framework. 
\textit{Nuclear Instruments and Methods in Physics Research Section A: Accelerators, 
Spectrometers, Detectors and Associated Equipment} \textbf{389} (1-2), 81-86 (1997).

\item[44.] Maragioglio, V. \& Rinaldi, C., \textit{L'architettura delle piramidi Menfite. 
Parte IV, La Grande piramide di Cheope} (Rapallo, 1965).

\item[45.] Kraus, K., \textit{Photogrammetry: geometry from images and laser scans} (Walter 
de Gruyter, 2007). 
\end{itemize}

\bigskip

\noindent\textbf{Extended Data Table 1 \textbar{} Comparison of the three muon detection 
technologies.}

\begin{center}
    \includegraphics{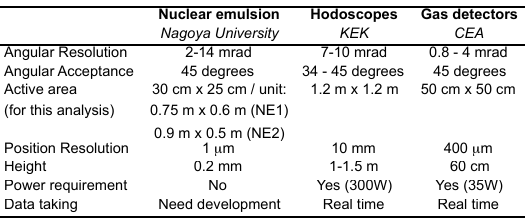}
\end{center}

\begin{SI-figure*}
    \includegraphics{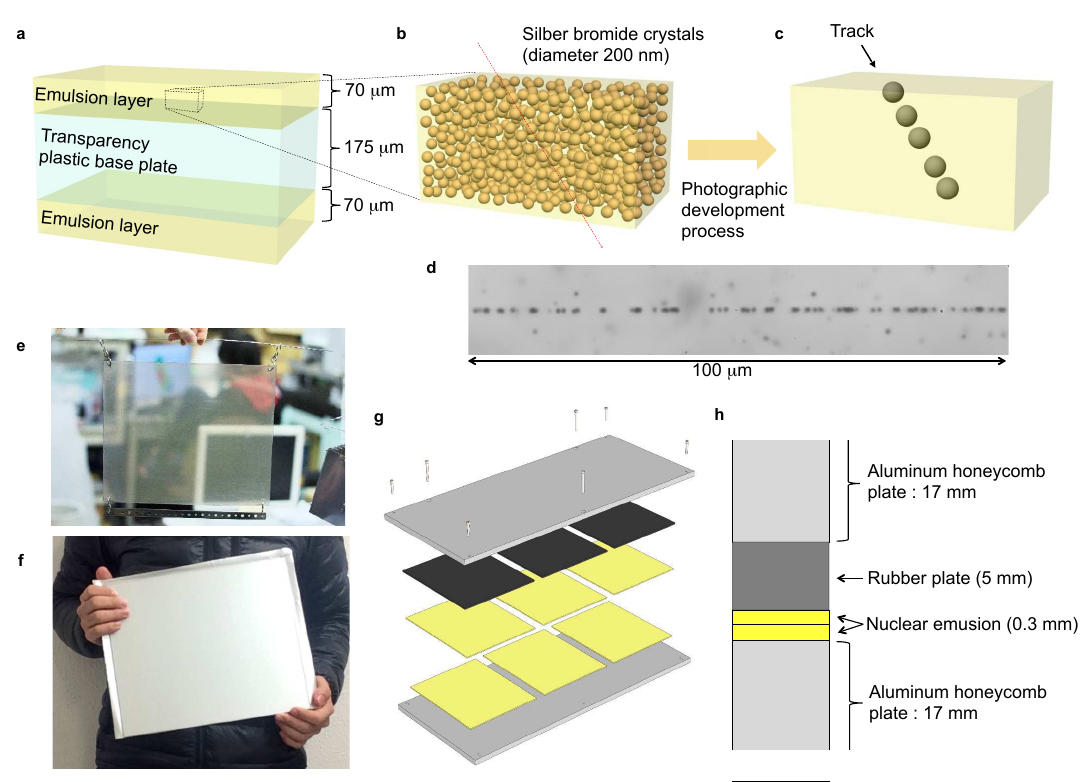}
\caption{\textbf{Overview of the nuclear emulsion films 
(Nagoya University).} \textbf{a}, A cross sectional schematic view of nuclear emulsion. 
\textbf{b}, Enlarged schematic view of the emulsion layer. Silver bromide crystals 
are dispersed in gelatin. The red dashed arrow shows the trajectory of the charged 
particle. \textbf{c}, After the photographic development process, silver grains 
are aligned along the lines with the trajectory (or, \textit{track}) of charged 
particle. \textbf{d}, An optical microscopic photograph of a track of minimum ionized 
particle recorded in a nuclear emulsion. \textbf{e}, A nuclear emulsion after development. 
\textbf{f}, A vacuum packed nuclear emulsion. \textbf{g}, Schematic view of the 
detector configuration: six packed nuclear emulsion films with a detection area 
of 30 × 25 cm\textsuperscript{2} each (yellow) are fixed between aluminum supporting 
plates (honeycomb plate, in grey). Two films stacked on top of each other are pressed 
with a rubber sheet (black) by four short screws. Three additional long screws 
are used as legs to correct the inclination of the detector. \textbf{h}, Cross 
sectional schematic view of the nuclear emulsion detector as shown in g. Two packed 
films are stacked between two honeycomb plates and rubber sheet.}
\end{SI-figure*}

\begin{SI-figure*}
    \begin{center}
    \includegraphics{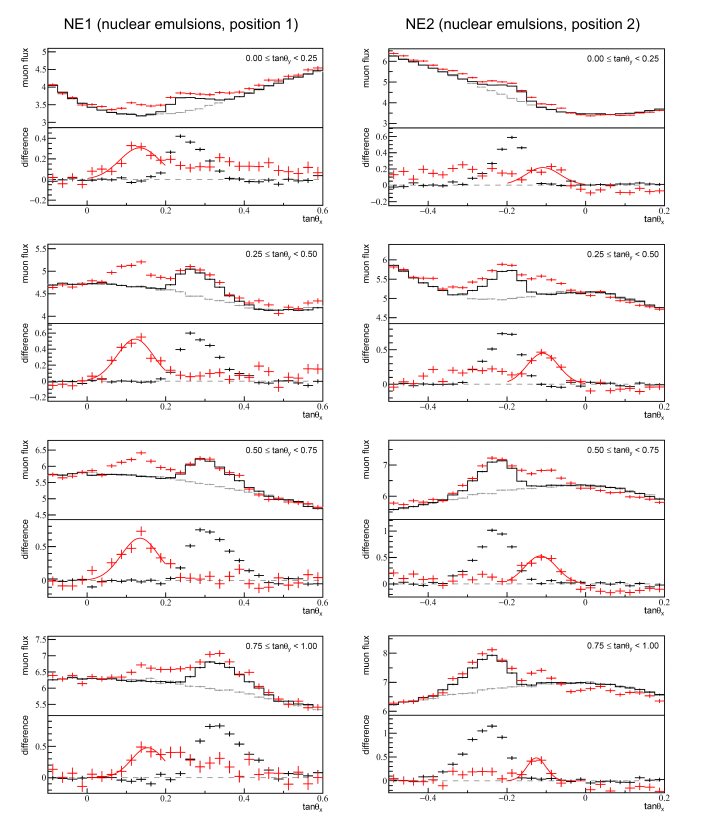}
    \end{center}
\caption{\textbf{Slices of the data for the nuclear 
emulsion plates.} Each figure show slices for tan$\theta$\textsubscript{y} every 
0.25 units (in tangent) and separated into four ranges at 0 $\leq$ tan$\theta$\textsubscript{y} 
\texttt{<} 1 (Fig. 2a,c). Top panel shows muon flux distribution and bottom panel 
shows difference of muon flux in each figure. In the top panels, the red line shows 
the data, the black solid line shows the simulation with the inner structures, 
and the grey dashed line shows the simulation without any internal structure. In 
the bottom panels, the red line shows the subtraction between the data and the 
simulation with the inner structures, the black line shows subtraction between 
the simulation with the inner structures and without them, so that the Grand Gallery 
appears as a muon excess. Error bars indicate 1 sigma of statistical error (standard 
deviation). The comparison between the excess that correspond to the Grand Gallery 
and the one that corresponds to the new void shows that the two structures are 
of a similar scale. For each projection of difference of muon flux, we performed 
a Gaussian fitting to estimate the direction of anomalies. The fitting zone was 
0 $\leq$ tan$\theta$\textsubscript{x} $\leq$ 0.2 for position NE1 and -0.2 $\leq$ tan$\theta$\textsubscript{x} 
$\leq$ 0 for position NE2. These fitted centres were used for the triangulation.}
\end{SI-figure*}

\begin{SI-figure*}
    \begin{center}
        \includegraphics{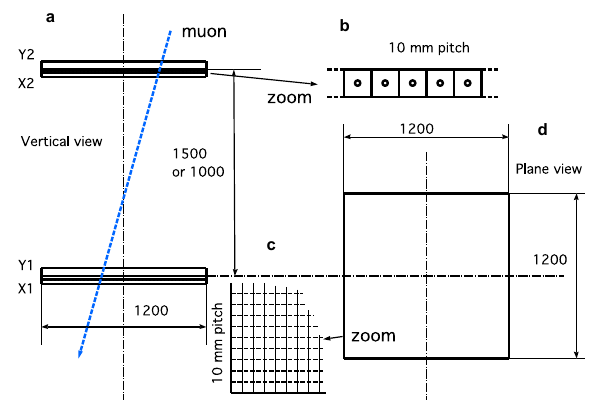}
    \end{center}
\caption{\textbf{Overview of the scintillator  hodoscopes (KEK)}. \textbf{a}, 
Vertical view of the detector, consisting of two 
units of orthogonal double scintillator layers. A blue arrow indicates a muon track 
passing through the whole instrument. \textbf{b}, Cross section of a scintillator 
element, showing the central hole for the optical fiber. \textbf{b}, Grid made 
of double layers, detecting the position of incident muons. \textbf{d}, Plane view 
of the detector having an active area of 1.2 × 1.2 m².}
\end{SI-figure*}

\begin{SI-figure*}
    \begin{center}
        \includegraphics{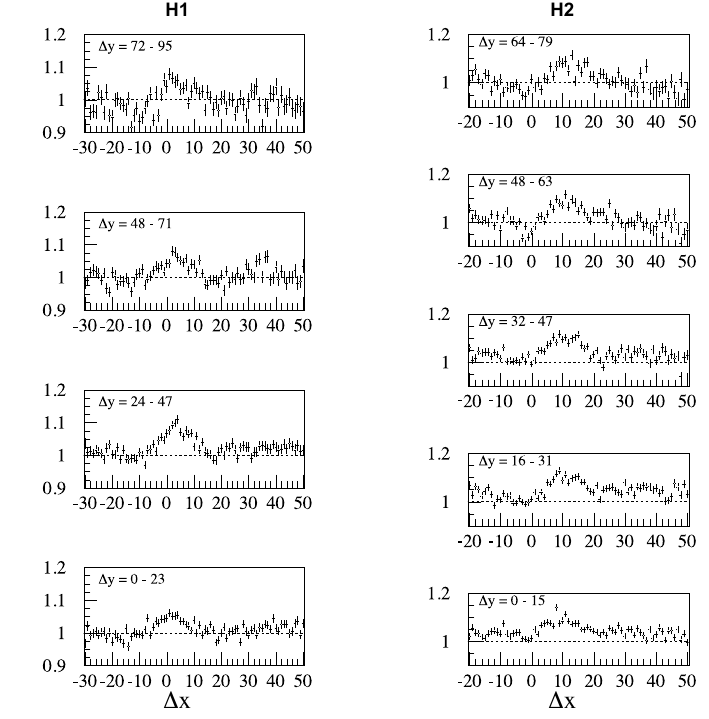}
    \end{center}
\caption{\textbf{Slices of the data for the scintillator 
hodoscopes. a,} Relative yield of the measurement to the simulation (including 
known structures) at position H1 for 4 slices (the width of each bin is 24 $\Delta$y). 
\textbf{b,} Relative yield at position H2 for 5 slices (the width of each bin is 
16 $\Delta$y). Error bars show one standard deviation.}
\end{SI-figure*}

\begin{SI-figure*}
    \begin{center}
        \includegraphics{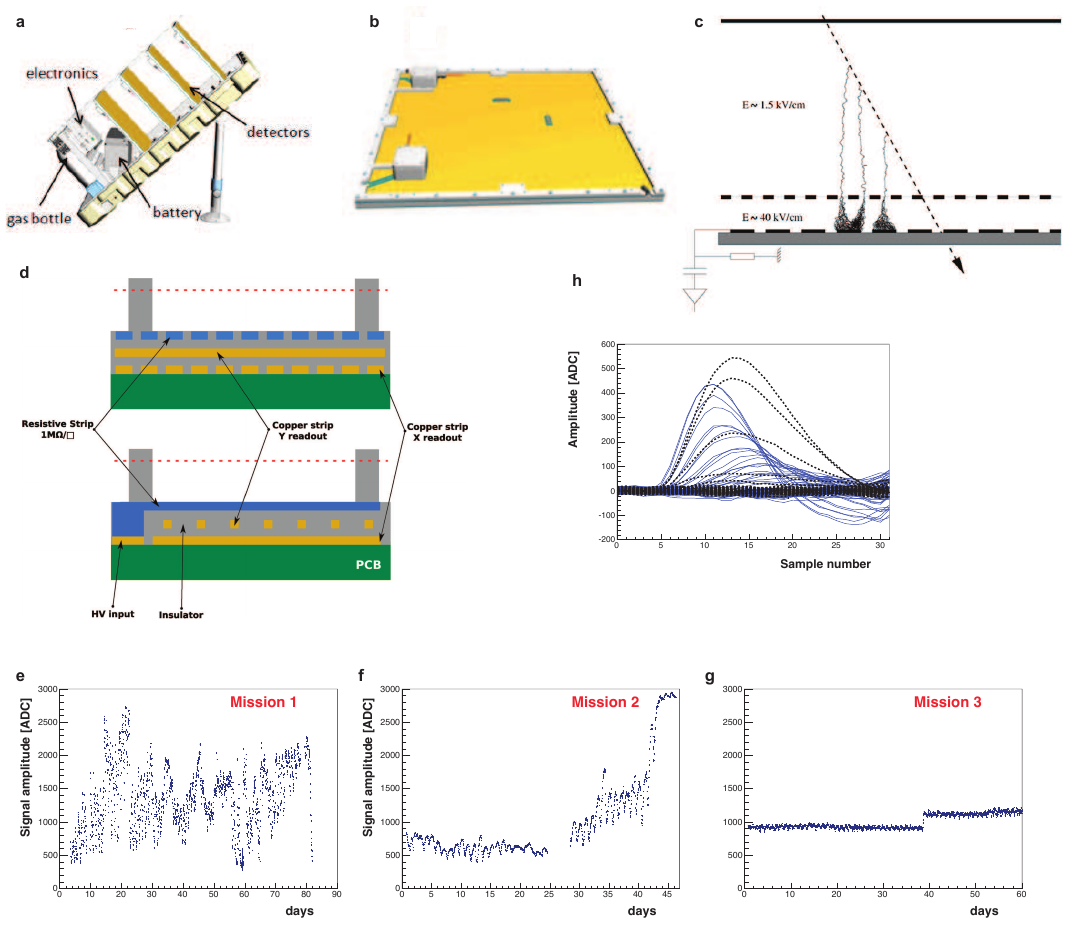}
    \end{center}
\caption{\textbf{Overview of the gas detectors 
(CEA). a,~}Design of a telescope (without its cover) showing the 4 detectors, the 
electronics box, the battery and the gas bottles. \textbf{b,} Design of the multiplexed 
Micromegas detector; \textbf{c,} Principle of a Micromegas detector showing the 
ionization and amplification of the signal initiated from a charged particle (dotted 
array). \textbf{d,} layout of the detector with the micro-mesh in red, the resistive 
strip film in blue, and the Y, X Copper readout strips in yellow; \textbf{e-g}, 
amplitude variation of a detector in Alhazen as a function of time for 2 previous 
campaigns (missions 1 and 2) and the one reported here (g), showing the effect 
of the patented feedback. Large variations observed in (e,f) can lead to inefficiency 
or degraded resolution, and are totally absent for the data of this paper. The 
only step observed in (g) corresponds to a manual change of the target amplitude. 
\textbf{h,} typical signal recorded in a detector, where each line corresponds 
to an electronic channel.}
\end{SI-figure*}

\begin{SI-figure*}
    \begin{center}
        \includegraphics{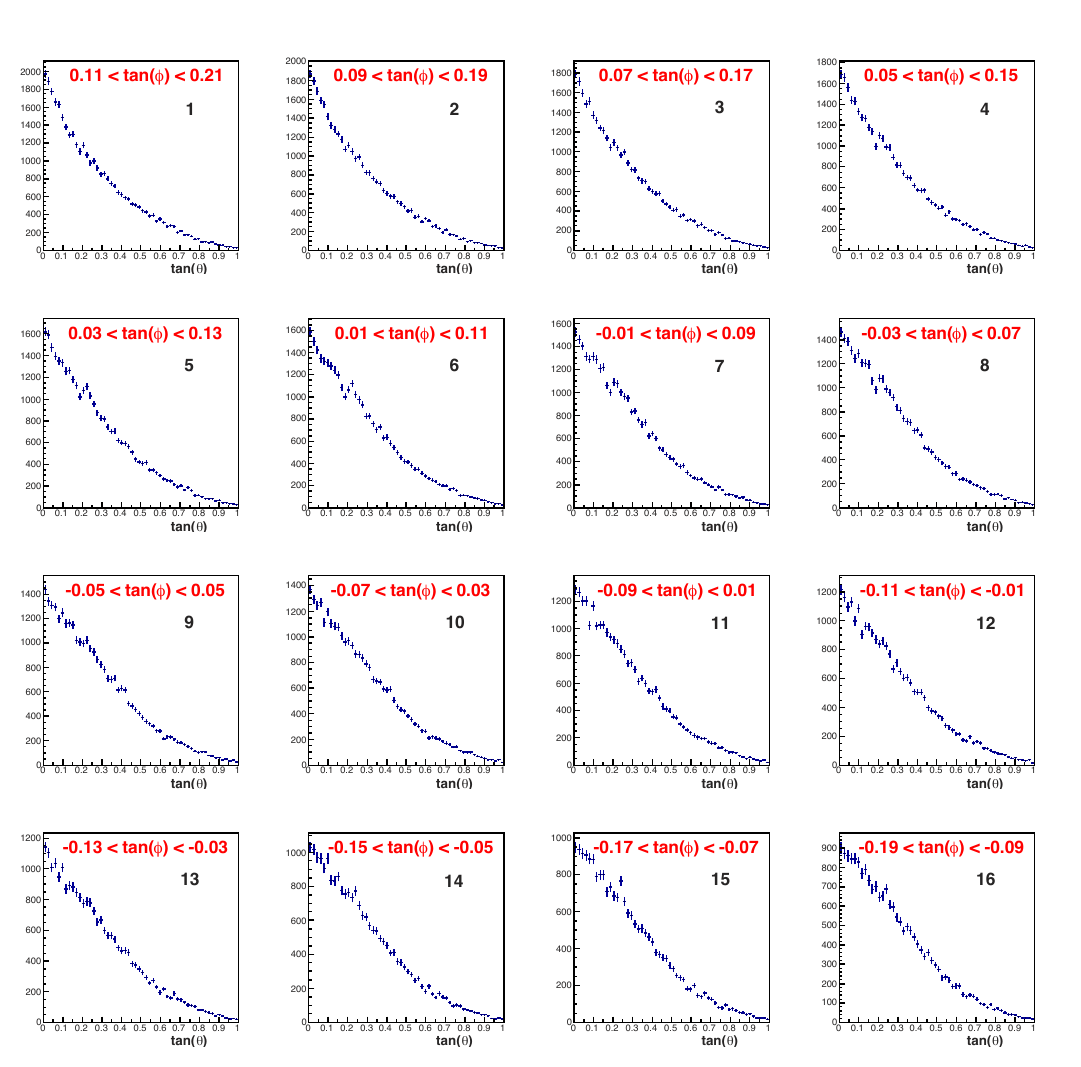}
    \end{center}
\caption{\textbf{tan(phi), horizontal slices 
on the Alhazen muography.} The slices are 0.10 (in tangent) thick, and each slice 
is shifted by 0.02 with respect to the previous one, which means they overlap. 
Distributions are generally smooth, with 2 significant muon excesses on histograms 
5-6 and 15 (see Methods for details). Error bars show one standard deviation.}
\end{SI-figure*}

\begin{SI-figure*}
    \begin{center}
        \includegraphics{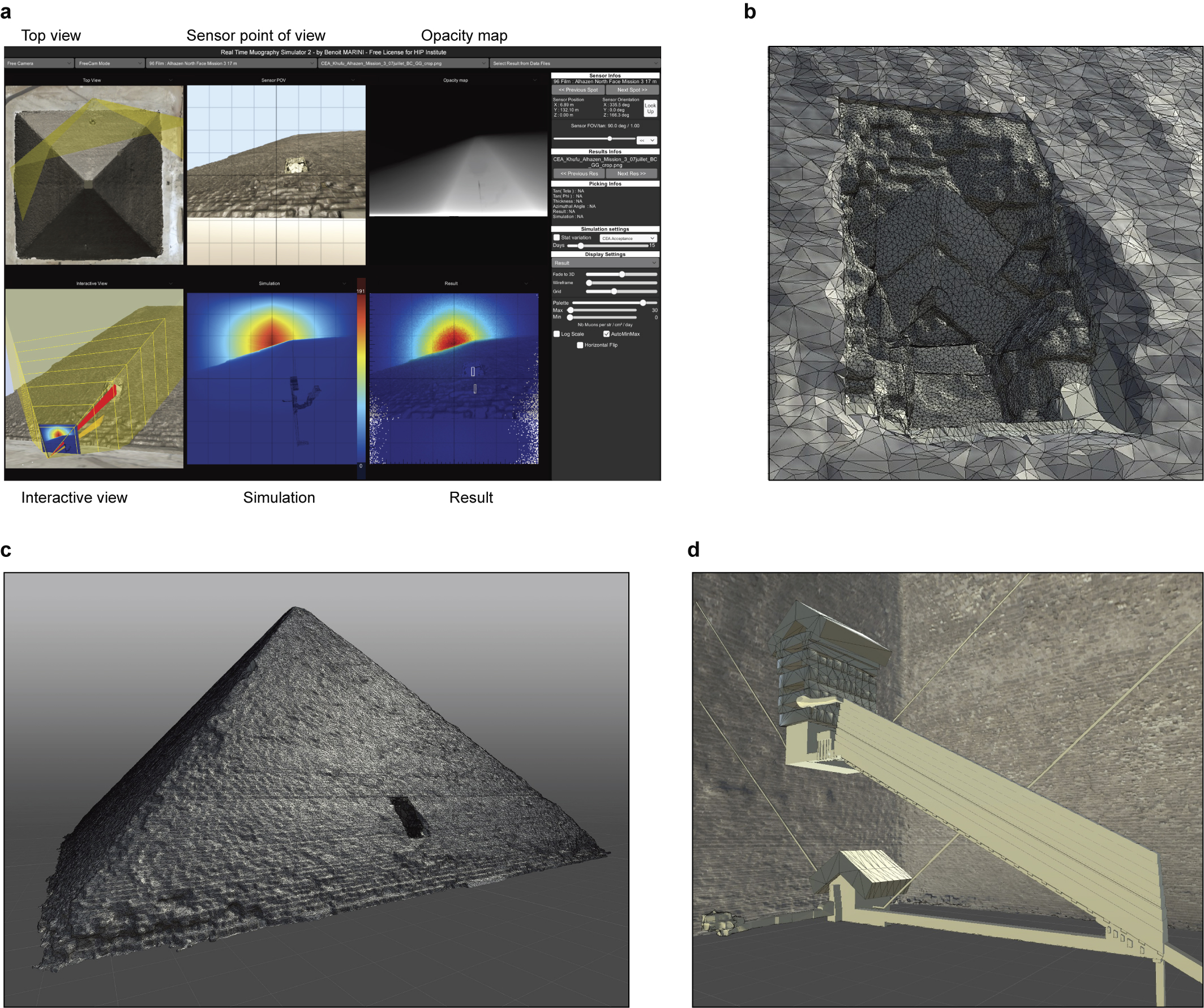}
    \end{center}
    \caption{\textbf{RTMS simulator and 3D models.} 
\textbf{a},~RTMS with CEA position G2 (Alhazen) sensor with a 6-view layout. Interactive 
view with display of sensor field of view and cone projections. Real time simulation 
with internal structure overlaid in wireframe. Result with sensor point of view 
superposition~\textbf{b}, Zoom on chevron area (shaded wireframe)~\textbf{c}, Large 
view of optimized 3D model (shaded wireframe).~\textbf{d}, Detail of optimized 
3D model.}
\end{SI-figure*}

\end{document}